\documentclass[aps,twocolumn,pra,superscriptaddress,amsmath,showpacs,tightenlines]{revtex4}
\usepackage{epsfig,graphicx,times}
\usepackage{amstext}
\usepackage{amsmath}            
\usepackage{amssymb}            
\usepackage{graphicx}           
\usepackage{latexsym}
\usepackage{bm}

\begin{document}

\title{Optomechanical analog of two-color electromagnetically-induced
transparency: Photon transmission through an optomechanical device
with a two-level system}

\author{Hui Wang}
\affiliation{Institute of Microelectronics, Tsinghua University,
Beijing 100084, China} \affiliation{CEMS, RIKEN, Saitama 351-0198,
Japan}

\author{Xiu Gu}
\affiliation{Institute of Microelectronics, Tsinghua University,
Beijing 100084, China}

\author{Yu-xi Liu}
\email{yuxiliu@mail.tsinghua.edu.cn} \affiliation{Institute of
Microelectronics, Tsinghua University, Beijing 100084, China}
\affiliation{Tsinghua National Laboratory for Information Science
and Technology (TNList), Beijing 100084, China} \affiliation{CEMS,
RIKEN, Saitama 351-0198, Japan}

\author{Adam Miranowicz}
\affiliation{Faculty of Physics, Adam Mickiewicz University,
61-614 Pozna\'n, Poland} \affiliation{CEMS, RIKEN, Saitama
351-0198, Japan}

\author{Franco Nori}
\affiliation{CEMS, RIKEN, Saitama 351-0198, Japan}
\affiliation{Physics Department, The University of Michigan, Ann
Arbor, Michigan 48109-1040, USA}

\date{\today}



\begin{abstract}
Some optomechanical systems can be transparent to a probe field
when a strong driving field is applied. These systems can provide
an optomechanical analogue of electromagnetically-induced
transparency (EIT). We study the transmission of a probe field
through a hybrid optomechanical system consisting of a cavity and
a mechanical resonator with a two-level system (qubit). The qubit
might be an intrinsic defect inside the mechanical resonator, a
superconducting artificial atom, or another two-level system. The
mechanical resonator is coupled to the cavity field via radiation
pressure and to the qubit via the Jaynes-Cummings interaction. We
find that the dressed two-level system and mechanical phonon can
form two sets of three-level systems. Thus, there are two
transparency windows in the discussed system. We interpret this
effect as an optomechanical analog of two-color EIT (or
double-EIT). We demonstrate how to switch between one and two EIT
windows by changing the transition frequency of the qubit. We show
that the absorption and dispersion of the system are mainly
affected by the qubit-phonon coupling strength and the transition
frequency of the qubit.
\end{abstract}

\maketitle

\section{Introduction}

Micro- and nano-scale mechanical resonators~\cite{PR1,PR2} provide
a platform to explore the transition from quantum physics to
classical physics. {Such transition can be demonstrated by
coupling  mechanical resonators to other quantum
objects~\cite{ZLXiang}, including superconducting qubit
circuits~\cite{Irish,amour,Sornborger,Zhang,Xue,cleland,tianT,wei-liu,yuxiliu,Roukes1,feixue1},
transmission line resonators~\cite{feixue2,yongli,T1,T2,T3},
optical cavities~\cite{review1,review2,review3,review4},
nitrogen-vacancy (NV) centers~\cite{NV1,NV2,NV3}, electron
spin~\cite{Rugar}, and two-level
defects~\cite{Tdefects,Grabovskij}. For example, the quantization
of mechanical oscillations can be demonstrated by phonon
blockade~\cite{yuxiliu}, which can be measured by a cavity
field~\cite{phononblockade}.} Experiments~\cite{Q1,Q2,Q3,Q4}
showed that mechanical resonators can be operated in the quantum
regime. This makes it possible to couple different degrees of
freedom in hybrid quantum devices~\cite{ZLXiang} using mechanical
resonators as quantum transducers~\cite{sun, Pirkkalainen},
switches, or data buses~\cite{gaoming}.

{It is well known that optomechanical
systems~\cite{review1,review2,review3,review4} can be created when
a mechanical resonator is coupled to electromagnetic fields
through radiation pressure. Although the mechanical resonator can
be coupled to electromagnetic fields at very different
wavelengths, most recent experiments use optomechanical couplings
from microwave to optical wavelengths. It has been shown both
theoretically~\cite{tian1,ying-dan,tian} and experimentally
~\cite{Hill,Hailing,cleland2013} that mechanical resonators can be
used to convert optical quantum states to microwave ones via
optomechanical interactions between a mechanical resonator and a
single-mode field of both optical and microwave wavelengths.
Hybrid electro-optomechanical systems can exhibit controllable
strong Kerr nonlinearities even in the weak-coupling
regime~\cite{Lu13}. This Kerr nonlinearity can enable, in
particular, the appearance of photon blockade or the generation of
nonclassical states of microwave radiation (e.g.,
two-component~\cite{cat} and multi-component~\cite{kitten}
Schr\"odinger cat states).} Also when a weak coherent probe field
is applied to the cavity of an optomechanical system, the
mechanical resonator can act as a switch to control the probe
photon transmission such that photons can pass through the cavity
one by one~\cite{blockade2,binghe,JQLiao1,JQLiao2} or two by
two~\cite{xunwei,germany} in the limit of the strong single-photon
optomechanical coupling~\cite{strong1,strong2,strong3,strong4}.
This phonon-induced photon blockade can be used to engineer
nonclassical phonon states~\cite{xu-phonon1,xu-phonon2} of
macroscopic mechanical resonators in low frequencies. Moreover,
optomechanical systems can also become transparent to a weak probe
field when a strong driving field is applied to the cavity field.
This is an optomechanical analogue of electromagnetically-induced
transparency (EIT)~\cite{Agarwal,Weis,SafaviNaeini,yingwu}.

{EIT was first observed in a Sr atom gas in 1991~\cite{Harris1}.
Sr atoms have a $\Lambda$-type three-level structure. The
destructive interference of two dressed states (resulting from the
strong coupling of the upper two energy levels) leads to an EIT
window~\cite{Harris2}. A similar energy-level structure also
exists in two coupled whispering-gallery-mode (WGM) microtoroidal
resonators~\cite{Peng}.  In an optomechanical system, analogously
to the Sr atom, a photon-energy level and the corresponding two
first-order sideband energy levels can also constitute a
$\Lambda$-type three-level structure~\cite{Weis}. The destructive
interference in the photon and phonon transition processes  can
lead to EIT phenomena in optomechanical systems. This has been
studied both theoretically and experimentally in, e.g.,
Refs.~\cite{Agarwal,Weis}.}

{Experiments have shown that both a cavity field and a mechanical
resonator can be coupled to other systems, thus optomechanical
systems are important ingredients in quantum networks. Recently,
many theoretical works studied the optical properties of an
optomechanical system coupled to a two-level or three-level
systems through the cavity field. For example, the theoretical
study~\cite{Ian} showed that when a two-level atomic ensemble is
coupled to the cavity field of an optomechanical system, it can be
used to enhance the photon-phonon coupling through radiation
pressure. It was also found that the EIT in a three-level atomic
ensemble, interacting with a cavity field of an optomechanical
system, can be significantly changed by an oscillating
mirror~\cite{yuechang}. Furthermore, the optomechanical coupling
was studied in a system where a single-mode cavity field is
coupled to an antiferromagnetic Bose-Einstein
condensate~\cite{huijing,gaoxiangli}, where the mechanical element
is provided by spin-wave excitations. It has also been
theoretically shown that an optomechanical analogue of EIT can be
controlled by a tunable superconducting qubit~\cite{huiwang} or a
two-level atom~\cite{hybrid1,hybrid7}, which is coupled to a
cavity. Moreover,  EIT and the related Autler-Townes splitting
phenomena have also been studied in superconducting artificial
atomic systems (e.g., Refs.~\cite{Ian10,Sun13,Bsanders} and the
many references therein).}

{It is also well known that impurities or defects usually exist in
bulk crystals and mechanical resonators. Such defects can affect
the quantum properties of a mechanical
resonator~\cite{Phillips,Mohanty,Arcizet}. Usually, the defects
are modeled as two-level systems, and they can interact with the
mechanical mode through a deformation force. Then, if there exists
a defect in the mechanical resonator of an optomechanical system,
then a three-body hybrid system composed of a mechanical mode, a
two-level defect, and a cavity field can be formed. The intrinsic
two-level defects in the mechanical resonator can affect the
ground-state cooling of the mechanical mode and the nonlinear
properties of an optomechanical system~\cite{Tdefects,hybrid2}. As
mentioned above, this kind of hybrid systems can also be realized
by an extra capacitive coupling of a superconducting qubit to the
mechanical resonator of an optomechanical system~\cite{Q2}. A more
complex hybrid system can be realized if the two-level system
interacts with both the cavity field and mechanical mode of an
optomechanical system~\cite{Hammerer,Pirkkalainen}. Therefore, one
can raise the interesting question of how a two-level system
affects the photon transmission through an optomechanical device
in which the two-level system is coupled to the mechanical mode,
in contrast to the case where the two-level system is coupled to
the cavity mode.}

{Here we theoretically study a general hybrid model, consisting of
an optomechanical device and a two-level system or a two-level
defect, which is coupled to a mechanical resonator described by
the Jaynes-Cummings Hamiltonian. Our work is motivated by (i) the
studies showing that an optomechanical device can be coupled to a
two-level system through the cavity field of the device, (ii) the
experimental progress on the coupling between a mechanical
resonator and a two-level system, e.g., superconducting
qubits~\cite{Q1}, and (iii) the observations that defects might
exist in the mechanical resonator of an optomechanical system. Our
numerical calculations are mainly focused on resonant interactions
between the low-frequency mechanical resonator and the two-level
system. The latter might be an intrinsic defect inside the
mechanical resonator, a superconducting artificial
atom~\cite{You}, or another two-level system. For simplicity,
hereafter we just use a qubit or a two-level system to denote
those kinds of systems.}

A main result of our work is the observation of the optomechanical
analog of two-color EIT and the demonstration how this EIT can be
switched to the standard single-color EIT. We are not aware of any
other works on this effect in optomechanical systems.
Nevertheless, two-color EIT (or double EIT) has already been
discussed in some other systems~\cite{2colorEIT}, e.g., in an
ensemble of two-level atoms coupled to a probe light or,
equivalently, a system of two-mode polaritons coupled to one
transition of the $\Lambda$-type three-level atoms~\cite{He07}.
This effect is also closely related to the EIT in a
double-$\Lambda$ system. Applications of two-color EIT include
nonlinear wave-mixing, cross-phase modulation, optical switching,
wavelength conversion, etc.

The paper is organized as follows: In Sec.~\ref{model}, we
describe a theoretical model and the equations of motion for the
system operators. In Sec.~\ref{steadystate}, we obtain
steady-state solutions of the system operators and further study
the stability of the system. In Sec.~\ref{EIT}, the light
transmission in this hybrid system is studied through the
input-output theory. In particular, the optomechanical analogues
of EIT are discussed here. We finally summarize our results in
Sec.~\ref{conclusion}.

\begin{figure}
\includegraphics[bb=130 180 540 510, width=8.8 cm, clip]{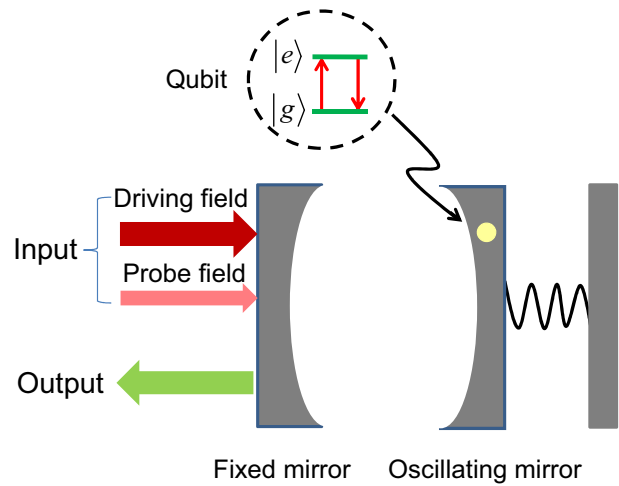}
\caption{(Color online) Schematic diagram of a hybrid
optomechanical system consisting of a cavity (with a photonic mode
$a$), where one of the mirrors is oscillating, corresponding to a
quantum mechanical resonator (with a phononic mode $b$). The
oscillating mirror has a qubit or two-level defect denoted by two
lines with the ground $|g\rangle$ and excited $|e\rangle$ states
inside the dashed circle. The mechanical resonator is coupled both
to the cavity (via radiation pressure) and to the qubit (via the
Jaynes-Cummings interaction). However, for simplicity, we assume
that there is no direct coupling between the cavity and
qubit.}\label{fig1}
\end{figure}

\section{Theoretical model}\label{model}

\subsection{Hamiltonian}

{As schematically shown in Fig.~\ref{fig1}, we study a general
theoretical model of a hybrid device in which a two-level system
is coupled to the mechanical resonator of an optomechanical
system~\cite{hybrid2}.} Figure~\ref{fig1} can also describe a
system, in which a single-mode cavity field is coupled to a
mechanical resonator which interacts with a two-level system as in
Refs.~\cite{amour,Q1}. We assume that the coupling between the
mechanical resonator and the two-level system is described by the
Jaynes-Cummings Hamiltonian. However, the interaction between the
mechanical resonator and the cavity field is described by the
radiation-pressure Hamiltonian. There is no direct interaction
between the two-level system and the cavity field. Thus, the
Hamiltonian of the whole system can be written as
\begin{eqnarray}\label{eq:1}
H_{0}&=&\hbar\omega_{a}a^{\dag}a+\hbar\omega_{b}b^{\dagger}b
+\frac{\hbar }{2}\omega_{q}\sigma_{z}-\hbar\chi a^{\dag}a \left(b^{\dagger}+b\right)\nonumber\\
&&+\hbar g\left(b^{\dagger}\sigma_{-} +\sigma_{+}b\right),
\end{eqnarray}
where $a$ ($a^{\dagger}$) is the annihilation (creation) operator
of the single-mode cavity field with frequency $\omega_{a}$; $b$
($b^{\dagger}$) is the annihilation (creation) operator of the
mechanical mode with frequency $\omega_{b}$. The  Pauli operator
$\sigma_{z}$  is used to describe the two-level system the
transition frequency $\omega_{q}$, while $\sigma_{+}$ and
$\sigma_{-}$ are the ladder operators of the two-level system. The
parameter $\chi$ is the coupling strength between the mechanical
resonator and the cavity field, while the parameter $g$ is the
coupling strength between the mechanical resonator and the
two-level system.

To demonstrate the relation between the cavity field and the
two-level system, let us apply a unitary transform $U=\exp[-\chi
a^{\dagger}a(b^{\dagger}-b)/\omega_{b}]$ to the Hamiltonian in
Eq.~(\ref{eq:1}). In this case, we have an effective Hamiltonian
$H^{\prime}_{0}=UH_{0}U^{\dagger}$ with
\begin{eqnarray}\label{eq:2}
H^{\prime}_{0}&=&\hbar\left(\omega_{a}-\frac{\chi^2}{\omega_{b}}+\frac{
g\chi}{\omega_{b}}\sigma_{x}\right)a^{\dag}a-\hbar
\frac{\chi^{2}}{\omega_{b}} a^{\dagger}a^{\dagger}aa
\nonumber\\
& &+\frac{\hbar
}{2}\omega_{q}\sigma_{z}+\hbar\omega_{b}b^{\dagger}b+\hbar
g\left(b^{\dagger}\sigma_{-} +\sigma_{+}b\right),
\end{eqnarray}
which shows that both the two-level system and the mechanical
resonator can induce a nonlinearity in the cavity field.

Let us now assume that a strong driving field and a weak probe
field, with frequencies $\omega_{d}$ and $\omega_{p}$,
respectively, are applied to the cavity. Then the Hamiltonian of
the driven hybrid system can be written as
\begin{eqnarray}\label{eq:3}
H&=&H_{0}+i\hbar\left(\Omega e^{-i\omega_{d}t}a^{\dag}-\Omega^{\ast} e^{i\omega_{d}t}a\right)\nonumber\\
&&+ i\hbar\left(\varepsilon
e^{-i\omega_{p}t}a^{\dag}-\varepsilon^{\ast}
e^{i\omega_{p}t}a\right),
\end{eqnarray}
where the parameters $\Omega$ and $\varepsilon$, with $|\Omega|\gg
|\varepsilon|$, correspond to the Rabi frequencies of the driving
(i.e., pump) and probe fields, respectively. In the rotating
reference frame with frequency $\omega_{d}$, the Hamiltonian in
Eq.~(\ref{eq:3}) becomes
\begin{eqnarray}\label{eq:4}
H_{r}&=&H_{0}-\hbar\omega_{d} a^{\dag}a+i\hbar\left(\Omega a^{\dag}-\Omega^{\ast}a\right)\nonumber\\
&&+i\hbar\left(\varepsilon e^{-i\Delta
t}a^{\dag}-\varepsilon^{\ast} e^{i\Delta t}a\right),\qquad
\end{eqnarray}
with the detuning $\Delta=\omega_{p}-\omega_{d}$ between the probe
field with frequency $\omega_{p}$ and the strong driving field
with frequency $\omega_{d}$.

\subsection{Heisenberg-Langevin equations}

Introducing the dissipation and fluctuation terms, and also using
the Markov approximation, the Heisenberg-Langevin equations of
motion can be written as
\begin{eqnarray}
\dot{a}&=&-(\gamma_{a}+i\Delta_{a})a+\Omega+\varepsilon \exp(-i\Delta t)+i\chi a\left( b^{\dagger}+ b\right)\nonumber\\
& &+\sqrt{2\gamma_{a}}a_{\rm{in}}(t),\label{eq:5}\quad\\
\dot{ b}&=&-\left(\gamma_{b}+i\omega_{b}\right) b+i\chi  a^{\dagger} a-i g \sigma_{-}+\sqrt{2\gamma_{b}}b_{\rm{in}}(t),\label{eq:6}\quad\\
\dot{\sigma}_{-}&=&-\left(\frac{\gamma_{q}}{2}+i\omega_{q}\right)\sigma_{-}+ig  b \sigma_{z}+\sqrt{\gamma_{q}}\,\Gamma_{-}(t), \label{eq:7}\\
 \dot{\sigma}_{z} &=&-\gamma_{q}\left( \sigma_{z} +1\right)-2i g\left(  b \sigma_{+}- b^{\dag} \sigma_{-} \right)+\sqrt{\gamma_{q}}\,\Gamma_{z}(t).\label{eq:8}
\end{eqnarray}
Here $\gamma_{a}$, $\gamma_{b}$, and $\gamma_{q}$ are the decay
rates of the cavity field, mechanical mode, and two-level system,
respectively. The parameter $\Delta_a=\omega_{a}-\omega_{d}$
describes the detuning between the cavity field $a$ with frequency
$\omega_{a}$ and the strong driving field with frequency
$\omega_{d}$. The operators $a_{\rm{in}}(t)$, $b_{\rm{in}}(t)$,
$\Gamma_{-}(t)$, and $\Gamma_{z}(t)$ denote environmental noises
corresponding to the operators $a$, $b$, $\sigma_{-}$, and
$\sigma_{z}$. We assume that the mean values of the above noise
operators are zero, that is,
\begin{equation}
\langle a_{\rm{in}}(t)\rangle=\langle
b_{\rm{in}}(t)\rangle=\langle \Gamma_{-}(t)\rangle=\langle
\Gamma_{z}(t)\rangle=0.
\end{equation}

\section{Steady states and stability analysis}\label{steadystate}

\begin{figure}
\includegraphics[bb=10 160 535 670, width=8.8 cm, clip]{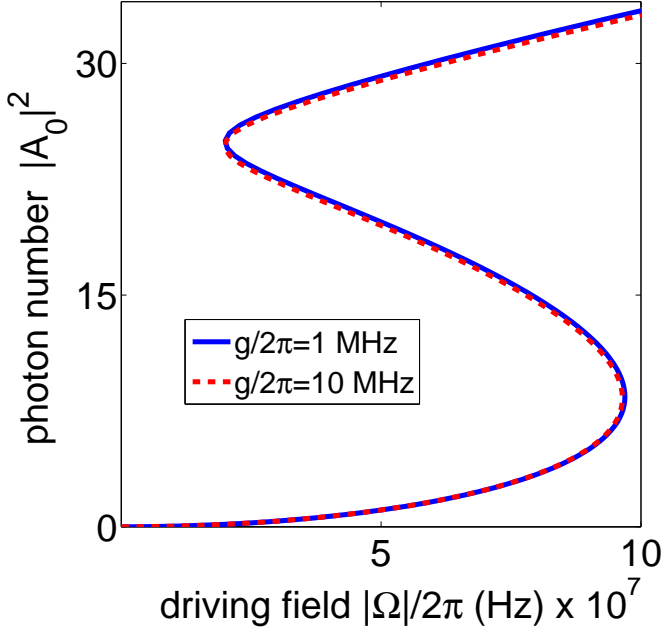}
\caption[]{(Color online) The steady-state photon number
$|A_{0}|^2$ of the cavity field is plotted as a function of the
driving field strength $|\Omega|/(2\pi)$, assuming $Z_{0}=-0.99$,
$g/(2\pi)=10 $ MHz, $\gamma_{a}/(2\pi)=4$ MHz,
$\omega_{q}/(2\pi)=\omega_{b}/(2\pi)=100$ MHz,
$\gamma_{q}/(2\pi)=0.1$ MHz, $\gamma_{b}/(2\pi)=1000$ Hz,
$\Delta_{a}/(2\pi)=50$ MHz, and $\chi/(2\pi)=10$ MHz. Note that
the frequency $\omega_{b}/(2\pi)$ of the mechanical mode is set to
$100$ MHz in the numerical calculations shown in all figures. }
\label{fig2}
\end{figure}

\subsection{Steady states and linear response to probe field}

To analyze the response of the system, in a steady state, to the
weak probe field, we now take the mean values corresponding to
Eqs.~(\ref{eq:5})--(\ref{eq:8}). In this case,  we have the
following equations
\begin{eqnarray}
\langle\dot{a}\rangle&=&-(\gamma_{a}+i\Delta_{a})\langle a\rangle+i\chi\langle a\rangle\left(\langle b^{\dagger}\rangle+\langle b\rangle\right)+\Omega\nonumber\\
& &+ \varepsilon \exp(-i\Delta t),\label{eq:15}\\
\langle\dot{ b}\rangle&=&-\left(\gamma_{b}+i\omega_{b}\right)\langle b\rangle+i\chi \langle a^{\dagger}\rangle\langle a\rangle-i g \langle\sigma_{-}\rangle,\label{eq:16}\\
\langle\dot{\sigma}_{-}\rangle&=&-\left(\frac{\gamma_{q}}{2}+i\omega_{q}\right)\langle\sigma_{-}\rangle+ig\langle
b\rangle \langle\sigma_{z}\rangle,\qquad\\ \label{eq:17}
\langle\dot{\sigma}_{z}\rangle&=&-\gamma_{q}\left(\langle\sigma_{z}\rangle+1\right)
-2i g\left(\langle b\rangle\langle\sigma_{+}\rangle-\langle
b^{\dag}\rangle\langle\sigma_{-}\rangle\right).\quad\label{eq:18}
\end{eqnarray}
Here we note that the mean-field approximation, i.e., $\langle
a^{\dagger}a\rangle=\langle a^{\dagger}\rangle \langle a\rangle $,
was used in the derivation of these equations.

It is very unlikely to obtain exact analytical solutions of the
nonlinear Eqs.~(\ref{eq:15})--(\ref{eq:18}), because the
steady-state response contains an infinite number of components of
different frequencies of the nonlinear systems. Instead, we find a
steady-state solution, which is exact for the driving field in the
parameter $\Omega$ and correct to first order in the parameter
$\varepsilon$ of the probe field. That is, we assume that the
solutions of Eqs.~(\ref{eq:15})--(\ref{eq:18}) have the following
forms~\cite{NonlinearOptics}:
\begin{eqnarray}
\langle a\rangle&=&A_{0}+A_{+}\exp(i\Delta t)+A_{-}\exp(-i\Delta t),\label{eq:19}\\
\langle b\rangle&=&B_{0}+B_{+}\exp(i\Delta t)+B_{-}\exp(-i\Delta t),\label{eq:20}\\
\langle \sigma_{-}\rangle&=&L_{0}+L_{+}\exp(i\Delta t)+L_{-}\exp(-i\Delta t),\label{eq:21}\\
\langle \sigma_{z}\rangle&=&Z_{0}+Z_{+}\exp(i\Delta t)
+Z_{-}\exp(-i\Delta t).\label{eq:22}
\end{eqnarray}
Here $A_{0} $,  $B_{0} $,  $L_{0}$, and $Z_{0}$ correspond to the
solutions of $a$, $b$, $ \sigma_{-}$, and $ \sigma_{z}$,
respectively, when only the driving field is applied. The
parameters $A_{\pm}$,  $B_{\pm}$, $L_{\pm}$, and $ Z_{\pm}$ are of
the order of $\varepsilon$ of the probe field. These can be
obtained by substituting Eqs.~(\ref{eq:19})--(\ref{eq:22}) into
Eqs.~(\ref{eq:15})--(\ref{eq:18}) and comparing the coefficients
of the same order. For example, we substitute the expressions
$\langle b\rangle $, $\langle \sigma_{-}\rangle $, and $\langle
\sigma_{z}\rangle$, given by Eqs.~(\ref{eq:20})--(\ref{eq:22}),
into Eq.~(\ref{eq:18}), then $Z_{0}$, $Z_{+}$, and $Z_{-}$ can be
expressed in terms of $L_{0}$, $L_{+}$, $L_{-}$, $B_{0}$, $B_{+}$,
and $B_{-}$ as follows
\begin{eqnarray}
Z_{0}&=& 2i \frac{g}{\gamma_{q}}\left( B^{\ast}_{0}L_{0}-B_{0}L^{\ast}_{0}\right)-1,\label{eq:23}\\
Z_{+}&=& -\lambda_{1}\left(B_{0}L^{\ast}_{-}+L^{\ast}_{0}B_{+}-B^{\ast}_{0}L_{+}-L_{0}B^{\ast}_{-}\right),\label{eq:24}\\
Z_{-}&=&
\lambda^{\ast}_{1}\left(B_{0}L^{\ast}_{+}+L^{\ast}_{0}B_{-}-B^{\ast}_{0}L_{-}-L_{0}B^{\ast}_{+}\right),\label{eq:25}
\end{eqnarray}
with $\lambda_{1}=2g/(\Delta-i\gamma_{q})$. Since $\sigma_{z}$ is
a Hermitian operator, the conditions $Z^{\ast}_{0}=Z_{0}$,
$Z^{\ast}_{+}=Z_{-}$, and $Z^{\ast}_{-}=Z_{+}$ have been used when
Eqs.~(\ref{eq:23})--(\ref{eq:25}) were derived. Similarly, we
obtain
\begin{eqnarray}
L_{0}=\frac{2 g B_{0}Z_{0}}{2\omega_{q}-i\gamma_{q}}.\label{eq:26}
\end{eqnarray}
Since the two-level system is coupled only to the mechanical mode,
then the steady-state value $L_{0}$ is directly related only to
the steady-state value of the mechanical mode and indirectly
related to those of the cavity field. We find that $L_{+}$ and
$L_{-}$ can be expressed with $B_{+}$ and $B_{-}$ as
\begin{eqnarray}
L_{+}&=&\lambda_{2}B_{+}+\lambda_{3}B^{\ast}_{-},\label{eq:27}\\
L_{-}&=&\lambda_{4}B_{-}+\lambda_{5}B^{\ast}_{+}.\label{eq:28}
\end{eqnarray}
Explicit formulas for these and other parameters $\lambda_{i}$
($i=2,3,...,10$) are given in the Appendix~\ref{App}.

We substitute $L_{0}$, given by Eq.~(\ref{eq:26}), into
Eq.~(\ref{eq:23}), and then obtain the solution
\begin{eqnarray}\label{eq:29}
Z_{0}=-\frac{\gamma^{2}_{q}+4\omega^{2}_{q}}{\gamma^{2}_{q}+4\omega^{2}_{q}+8g^{2}|B_{0}|^{2}}.
\end{eqnarray}
It is easy to find that the value of $Z_{0}$ ranges from $-1$ to
$0$. If $Z_{0}=-1$, then the two-level system is in its ground
state. Thus, it is obvious that if the coupling strength $g$ is
much smaller than the transition frequency $\omega_{q}$ of the
two-level system, and the phonon number is not very large, then
the two-level system is almost in its ground state. We also obtain
\begin{eqnarray}
B_{0}&=&\frac{\chi |A_{0}|^{2}- gL_{0}}{\omega_{b}-i\gamma_{b}},\label{eq:30}\\
B_{+}&=&\lambda_{6}(A^{\ast}_{0}A_{+}+A_{0}A^{\ast}_{-}),\label{eq:31}\\
B_{-}&=&\lambda_{7}(A^{\ast}_{0}A_{-}+A_{0}A^{\ast}_{+}),\label{eq:32}
\end{eqnarray}
where $B_{0}$ is the steady-state value of the mechanical mode.
Because of the coupling of the mechanical mode to the two-level
system and the cavity field, $B_{0}$ depends both on the
steady-state values $L_{0}$ of the two-level system and on the
steady-state value $A_{0}$ of the cavity field.

We now calculate the coefficients of the steady-state value of the
cavity field by substituting the expansions of $\langle a\rangle$,
given by Eq.~(\ref{eq:19}), and $\langle b\rangle$, given by
Eq.~(\ref{eq:20}), into Eq.~(\ref{eq:15}). Up to first order in
the parameter $\varepsilon$, we obtain
\begin{eqnarray}
A_{0}&=&\frac{\Omega}{\gamma_{a}+i\Delta_{a}-i\chi
\left(B_{0}+B^{\ast}_{0}\right)},\label{eq:33}
\end{eqnarray}
which represents the steady-state value of the cavity field
assuming a strong driving field. With the help of
Eqs.~(\ref{eq:31})--(\ref{eq:33}), we also obtain
\begin{eqnarray}
 A_{-}&=&\frac{\lambda_{9}\varepsilon}{\lambda_{8}\lambda_{9}-\lambda_{10} }, \label{eq:34}
\end{eqnarray}
which describes the response of the strongly-driven system to the
weak-probe field, when the driven system reaches a steady state.
Similarly, we also obtain
\begin{eqnarray}
 A_{+}&=&\frac{i\chi \left( \lambda_{6}+\lambda^{\ast}_{7}\right)A^{2}_{0}\varepsilon^{\ast}}{\lambda^{\ast}_{8}\lambda^{\ast}_{9}-\lambda^{\ast}_{10}},\label{eq:35}
\end{eqnarray}
which describes the four-wave mixing for the driving field and the
weak probe field.

\subsection{Stability}

To analyze   the stability of the system, we now present the
driving field strength $|\Omega|$ (i.e., the Rabi frequency of the
driving field) as a function of the steady value of $|A_{0}|$ as
follows
\begin{equation}
|\Omega|=|A_{0}|\sqrt{\gamma^{2}_{a}+\left[\Delta_{a}-\frac{2\chi^{2}\varepsilon_{2}
|A_{0}|^{2}
\left(\gamma^{2}_{a}+4\omega^{2}_{q}\right)}{\varepsilon^{2}_{1}+\varepsilon^{2}_{2}}\right]^{2}}\label{eq:40}
\end{equation}
with the parameters $\varepsilon_{1}$ and $\varepsilon_{2}$ given
by
\begin{eqnarray}
\varepsilon_{1}&=&\gamma_{b}\left(\gamma^{2}_{a}+4\omega^{2}_{q}\right)-2\gamma_{a}g^{2}Z_{0},\label{eq:36}\\
\varepsilon_{2}&=&\omega_{b}\left(\gamma^{2}_{a}+4\omega^{2}_{q}\right)+
4\omega_{q}g^{2}Z_{0}.\label{eq:37}
\end{eqnarray}
We can conclude  from Eq.~(\ref{eq:29}) that if the coupling
strength $g\ll \omega_{q}$ and the phonon number are small then
the two-level system has a high possibility to remain  in its
ground state. In this case the value of $Z_{0}$ is very close to
$-1$ and can be considered constant, then we can see from
Eq.~(\ref{eq:40}) that $|A_{0}|$ can have three real solutions
under certain conditions.

In Fig.~\ref{fig2}, the steady-state photon number $|A_{0}|^2$ of
the cavity field, corresponding to the steady-state component in
Eq.~(\ref{eq:19}), is plotted as a function of the driving field
strength $|\Omega|$. This figure shows the bistable behavior of
the cavity field of the hybrid optomechanical system. This result
is very similar to that of driven optomechanical
systems~\cite{mancini}. If we change the coupling strength between
the phonon and the two-level system, the steady value and bistable
behavior change.

The relation between the phonon mode $B_{0}$ and the Rabi
frequency $\Omega$ can also be calculated as
\begin{eqnarray}\label{eq:41}
|\Omega|^{2}&=&\frac{\left[\gamma^{2}_{a}+\left(\Delta_{a}-2\chi
{\rm Re}\,
B_{0}\right)^{2}\right]\left(\varepsilon_{4}+i\varepsilon_{5}\right)B_{0}}{i\chi\varepsilon_{3}}
\end{eqnarray}
with the parameters given by
\begin{eqnarray}
\varepsilon_{3}&=&\gamma^{2}_{a}+4\omega^{2}_{q}\\
\varepsilon_{4}&=&\gamma_{b}\varepsilon_{3}-2\gamma_{a}g^{2}Z_{0}\\
\varepsilon_{5}&=&\omega_{b}\varepsilon_{3}+ 4\omega_{q}g^{2}Z_{0}
\end{eqnarray}

\begin{figure}
\includegraphics[bb=30 165 530 680, width=8.8 cm, clip]{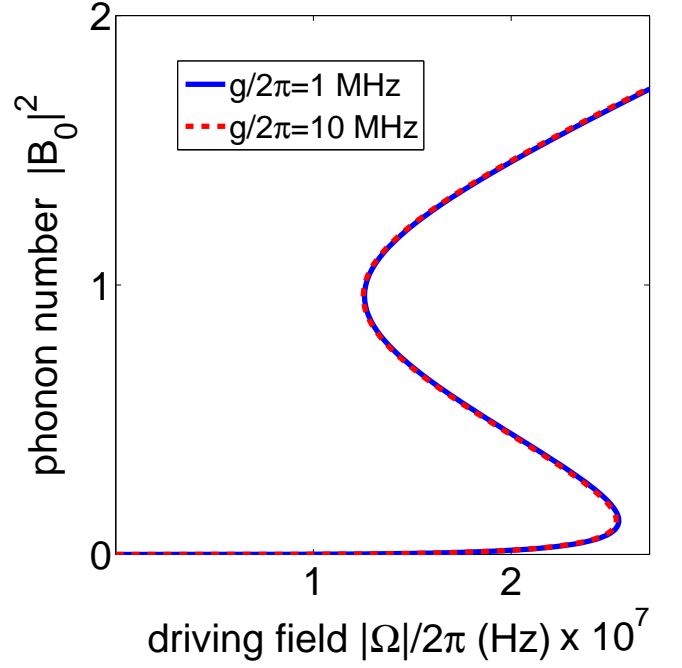}
\caption[]{(Color online)  The steady-state phonon number
$|B_{0}|^2$  of the mechanical mode is plotted as a function of
the driving field strength $|\Omega|/(2\pi)$ . The other
parameters are the same as in Fig.~\ref{fig2}, except that
$\Delta_{a}/(2\pi)=20$ MHz.}\label{fig3}
\end{figure}
In Fig.~\ref{fig3}, the steady-state phonon number $|B_{0}|^2$ of
the mechanical mode, corresponding to the steady-state component
in Eq.~(\ref{eq:20}), is plotted as a function of the Rabi
frequency $|\Omega|$ of the driving field. We find that  the
phonon bistability can also occur for the mechanical mode in some
parameter regimes, and the two-level system has a little effect on
the bistability. From Eqs.~(\ref{eq:40}) and (\ref{eq:41}), if the
variation of $Z_{0}$ cannot be ignored, we find that both
$|A_{0}|$ and $|B_{0}|$ can have at most five real solutions, so
both the photon and phonon modes can show multistability under
certain conditions. According to our numerical calculations shown
in Figs.~\ref{fig2} and~\ref{fig3}, if the coupling strength $g$
is much smaller than $\omega_{b}$ (or $\omega_{q}$),  the photon
and phonon modes only exhibit the bistable behavior, no
multistability. We note that multistability can occur when the
coupling between the two-level system and the phonon mode become
very strong or ultrastrong. This coupling might not be easy to
produce using natural qubits. However they might become possible
using an artificial two-level system, e.g., when the mechanical
mode is coupled to a superconducting qubit instead of an intrinsic
natural two-level defect.

\section{Electromagnetically-induced transparency}\label{EIT}

We now study the transmission of a weak-probe field through an
optomechanical system which is coupled to a two-level system.
Using the input-output theory~\cite{walls}
\begin{eqnarray}
\langle a_{\rm out}\rangle+\frac{\Omega}{\sqrt{2\gamma_{a}}}
+\frac{\varepsilon}{\sqrt{2\gamma_{a}}} e^{-i\Delta
t}=\sqrt{2\gamma_{a}}\langle a\rangle,\label{eq:42}
\end{eqnarray}
the output of the cavity field can be obtained as
\begin{eqnarray}\label{eq:43}
\langle a_{\rm {out}}\rangle &=&A_{d}+A_{s}\varepsilon e^{-i\Delta
t}+A_{as}\varepsilon^{\ast}e^{i\Delta  t},
\end{eqnarray}
with coefficients
\begin{eqnarray}
A_{d}&=&\sqrt{2\gamma_{a}}A_{0}-\frac{\Omega}{\sqrt{2\gamma_{a}}},\label{eq:44}\\
A_{s}&=&\frac{\sqrt{2\gamma_{a}}}{\varepsilon}A_{-}-\frac{1}{\sqrt{2\gamma_{a}}},\label{eq:45}\\
A_{as}&=&\frac{\sqrt{2\gamma_{a}}}{\varepsilon^{\ast}}A_{+}.\label{eq:46}
\end{eqnarray}
Here $A_{0}$ and $A_{\pm}$ are given by
Eqs.~(\ref{eq:33})--(\ref{eq:35}); $A_{d}$ is the output
responding to the driving (or control) field with frequency
$\omega_{d}$, $A_{s}$ is the output corresponding to the probe
field with frequency $\omega_{p}$ (Stokes frequency), and $A_{as}$
represents the four-wave mixing frequency $2\omega_{d}-\omega_{p}$
(anti-Stokes frequency). We redefine the output field at the
frequency $\omega_{p}$ of the probe field as
$\varepsilon_{T}=2\gamma_{a}A_{-}/\varepsilon$ with the real and
imaginary parts
$\mu_{p}=\gamma_{a}\left(A_{-}+A^{\ast}_{-}\right)/\varepsilon$
and
$\nu_{p}=\gamma_{a}\left(A_{-}-A^{\ast}_{-}\right)/(i\varepsilon)$.
It is clear that $\mu_{p} $ and $\nu_{p} $ describe absorption and
dispersion to the probe field.

For convenience, let us assume that
$\varepsilon=\sqrt{2\gamma_{a}P_{s}/\hbar\omega_{p}}$ is real.
Here $P_{s}$ is defined as the input power of the probe field.
Then the output power at the Stokes frequency relative to the
input power $P_{s}$ is
\begin{equation}
G_{s}=\frac{\hbar\omega_{p}|\varepsilon
A_{s}|^{2}}{P_{s}}=|\sqrt{2\gamma_{a}}A_{s}|^{2}.
\end{equation}
The output power at the anti-Stokes frequency
$2\omega_{d}-\omega_{p}$ is~\cite{Huang10}:
\begin{equation}
G_{as}=\frac{\hbar(2\omega_{d}-\omega_{p})|\varepsilon
A_{as}|^{2}}{P_{s}}=|\sqrt{2\gamma_{a}}A_{as}|^{2}.
\end{equation}

\begin{figure}
\includegraphics[bb=170 140 650 530, width=8.8 cm, clip]{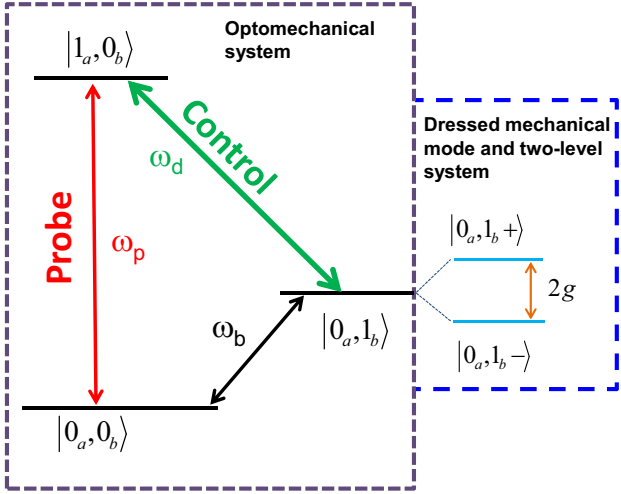}
\caption[]{(Color online) Schematic diagram for the interaction
between the hybrid system with the driving (control) and probe
fields with a single-particle excitation. The driving field is
applied to make a single-phonon transition from the phonon vacuum
state $|0_{b}\rangle$ to the single-phonon state $|1_{b}\rangle$.
The probe field is used to measure the transition when the
population of the mechanical mode is not changed. The
$\Lambda$-type three energy levels in the hybrid optomechanical
system occur in the case $\Delta_{a}=\omega_{b}$. Here
$|1_{b}\pm\rangle=(| 1_{b},g\rangle\pm |0_{b},e\rangle)/\sqrt{2}$
correspond to the dressed states between the single-phonon state
and the two-level system for $\omega_{b}=\omega_{q}$.}\label{fig4}
\end{figure}

In the resolved sideband limit $\omega_{b}\gg \gamma_{a}$, it is
known that the transmission spectrum exhibits an EIT analogue in
optomechanical systems. These phenomena can be mapped to the
$\Lambda$-type three-level diagram of atomic systems.

However, when the mechanical resonator of the optomechanical
system  is coupled to a two-level system, the transmission of the
probe field becomes complicated.  This is because the
Jaynes-Cummings coupling between the two-level system and the
mechanical resonator can lead to dressed states, which have a more
complicated energy structure. {This usually leads to a more
complicated
 absorption of a weak probe field by a hybrid system, and can be referred
  to as an EIT in a multi-level atom system~\cite{Shu1,Goren,Shu2}.}

\begin{figure}
 \includegraphics[bb=40 180 570 670, width=8.6 cm, clip]{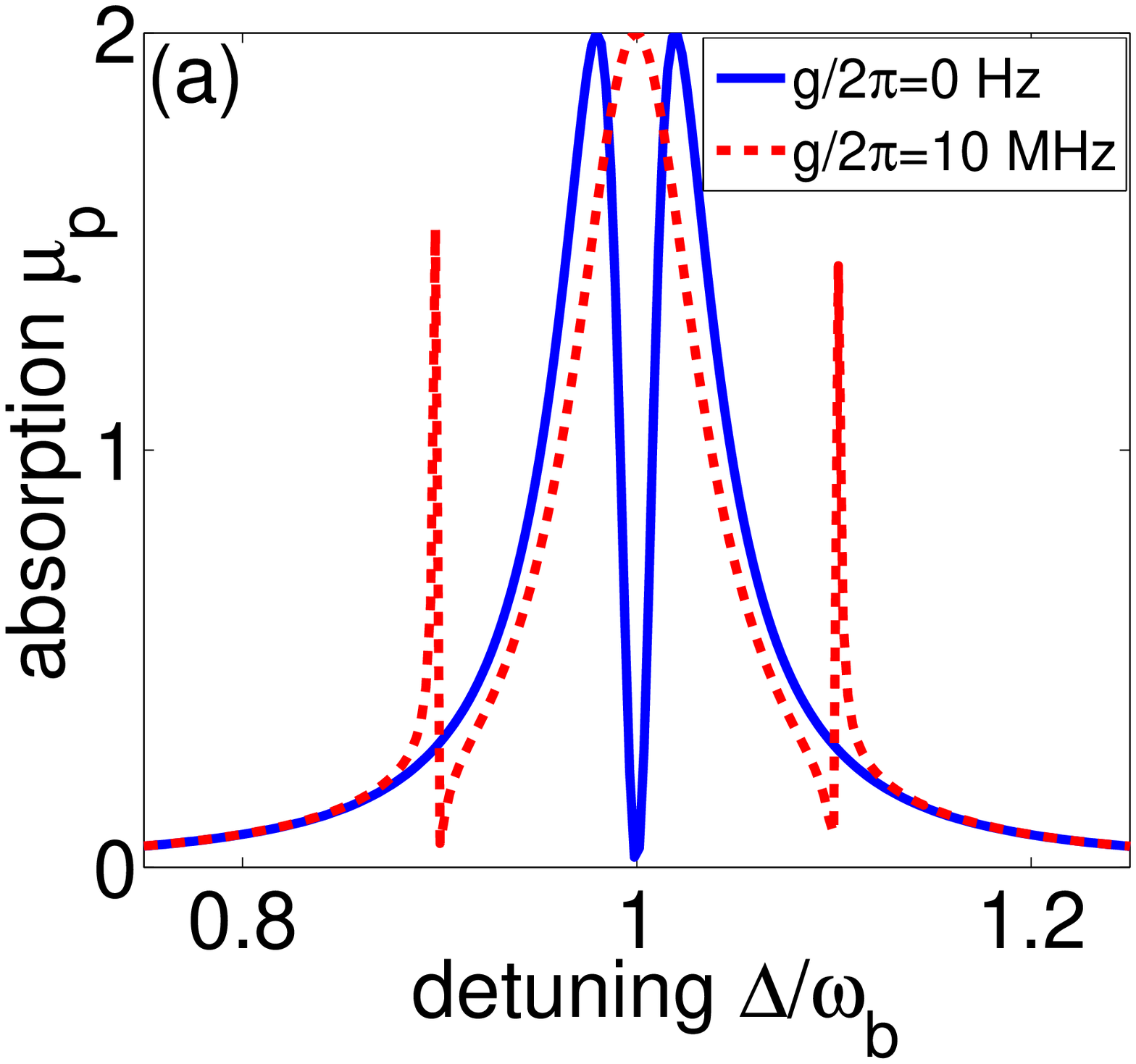}\\
 \includegraphics[bb=40 180 570 670, width=8.6 cm, clip]{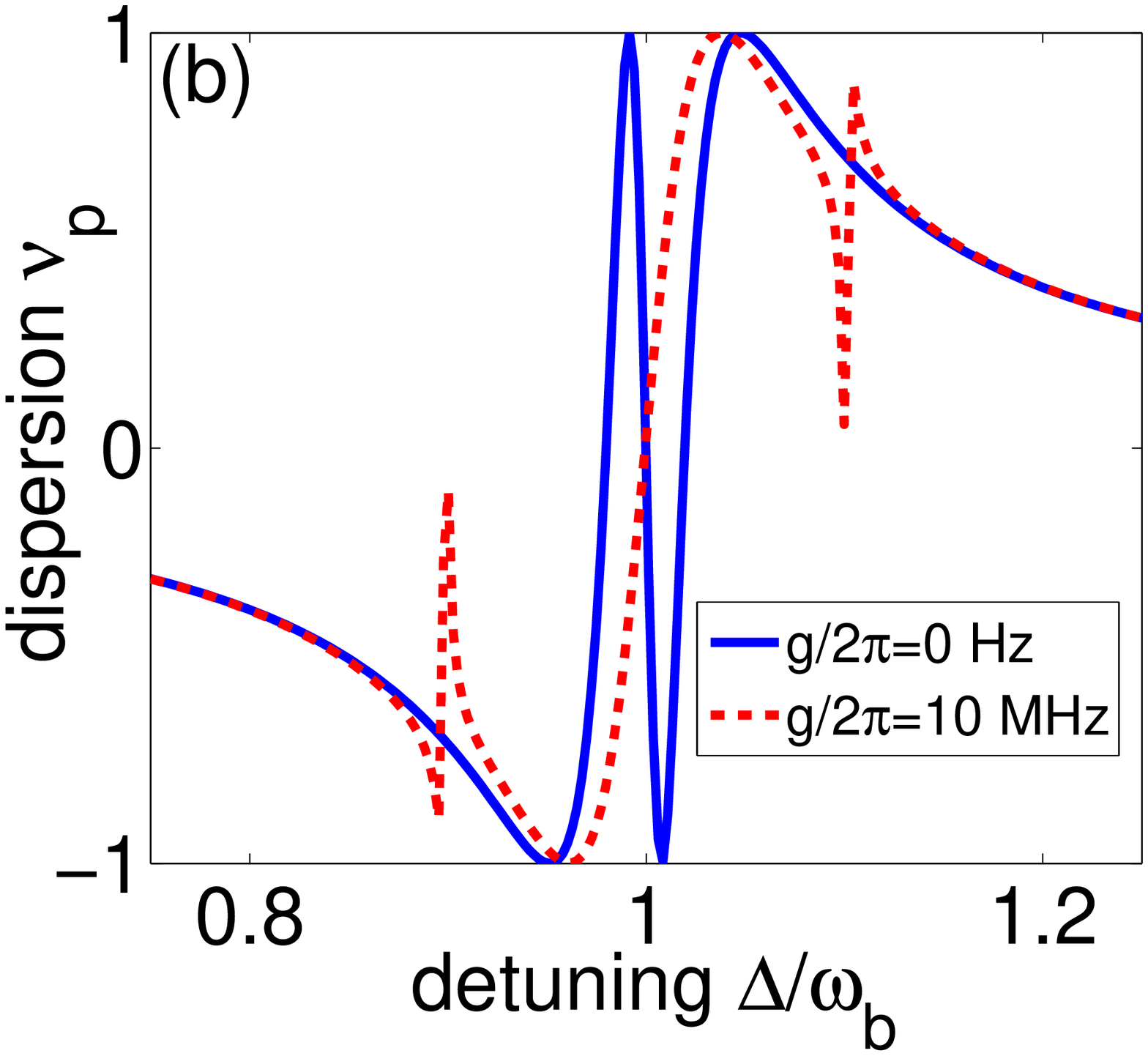}\\
\caption[]{(Color online) (a) Absorption $\mu_{p}$ and (b)
dispersion $\nu_{p}$ versus the detuning
$\Delta=\omega_{p}-\omega_{d}$ of the probe field in units of the
mechanical-mode frequency $\omega_{b}$. In each figure, the curves
correspond to different values [0 Hz (blue solid curve) and $10$
MHz (red dashed curve)] of the coupling strength $g/(2\pi)$
between the mechanical resonator and the two-level system. The
other parameters are the same as in Fig.~\ref{fig2}, except
$\Delta_{a}/(2\pi)=100$ MHz and $|\Omega|/(2\pi) =19.8$ MHz. The
blue and red curves describe the properties of the single
(standard) and double (or two-color) optomechanical EIT effects,
respectively. }\label{fig5}
\end{figure}

\begin{figure}
 \includegraphics[bb=10 150 550 630, width=8.6 cm, clip]{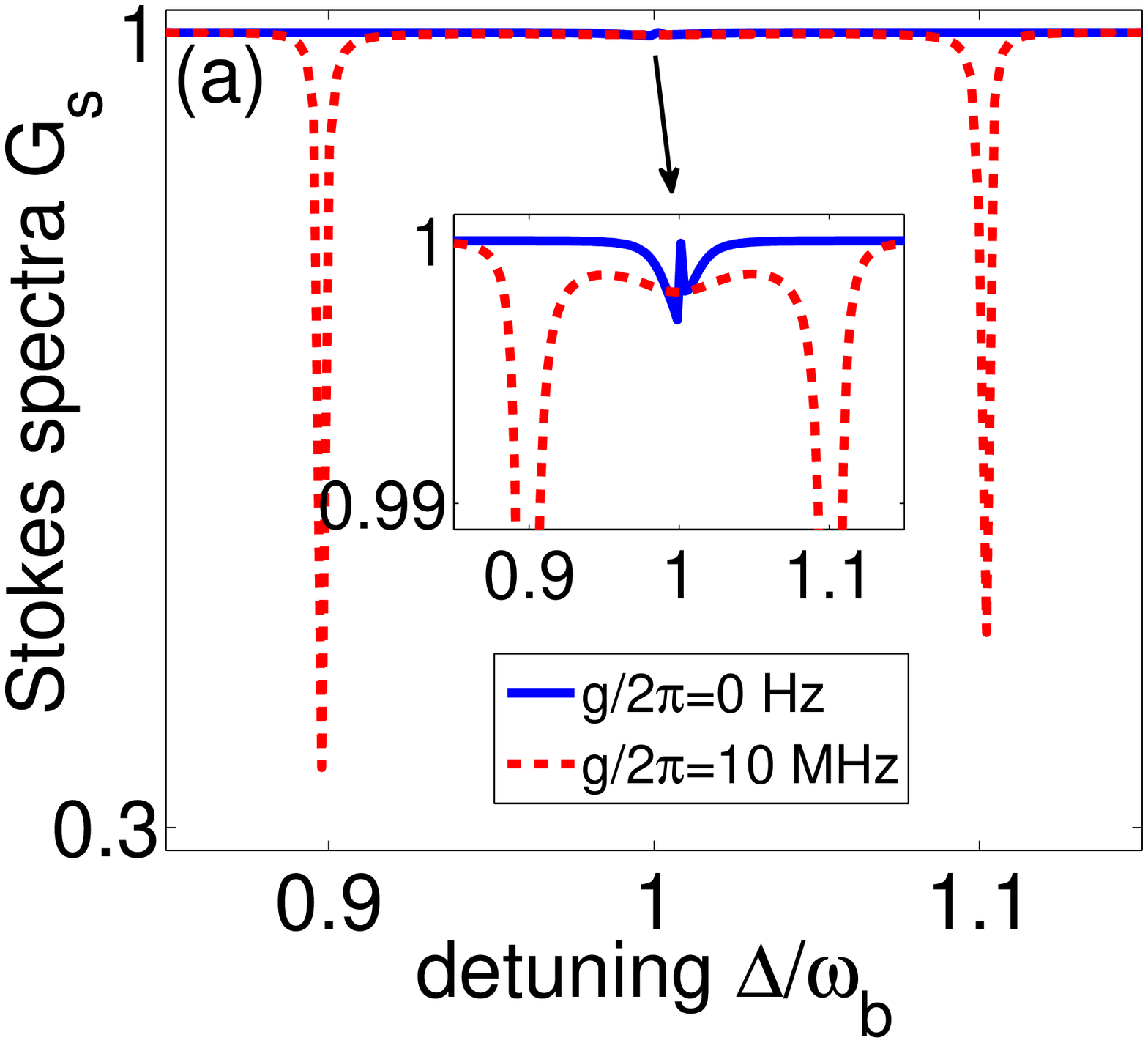}\\
 \includegraphics[bb=51 170 575 690, width=8.6 cm, clip]{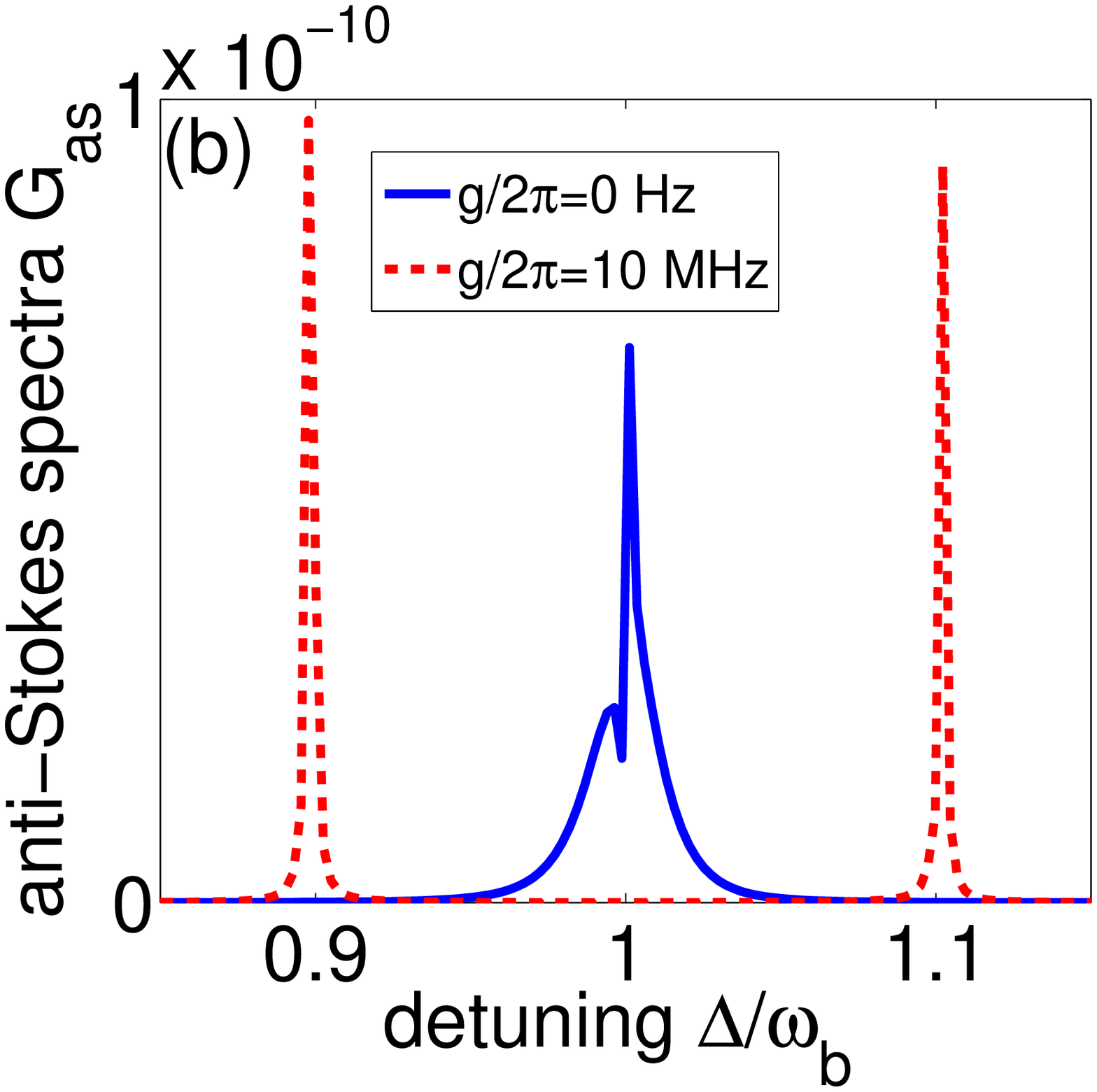}
\caption[]{(Color online) (a) The Stokes $G_{s}$ and (b)
anti-Stokes $G_{as}$ spectra of the output of the probe field
versus the detuning $\Delta=\omega_{p}-\omega_{d}$ of the probe
field in units of the mechanical-mode frequency $\omega_{b}$. The
parameters are the same as in Fig.~\ref{fig5}. }\label{fig6}
\end{figure}

\begin{figure}
 \includegraphics[bb=10 120 580 715, width=8.6 cm,clip]{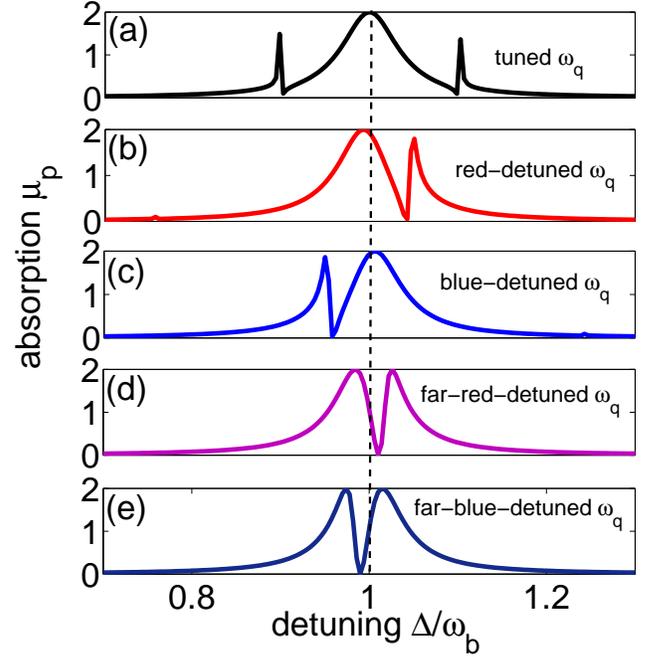}
\caption[]{(Color online) The absorption $\mu_p$ of the probe
field as a function of the detuning $\Delta=\omega_{p}-\omega_{d}$
between the probe frequency $\omega_{p}$ and the driving field
frequency $\omega_{d}$ for different values of the qubit
transition frequency: (a) $\omega_{q}/(2\pi)=100$ MHz, which
corresponds to the resonance of $\omega_{q}$ with the
mechanical-mode frequency $\omega_{b}$; (b) $\omega_{q}/(2\pi)=80$
MHz (i.e., $\omega_{q}<\omega_{b}$); (c) $\omega_{q}/(2\pi)=120$
MHz (i.e., $\omega_{q}>\omega_{b}$), (d) $\omega_{q}/(2\pi)=10$
MHz (i.e., $\omega_{b}-\omega_{q} \gg g$), and (e)
$\omega_{q}/(2\pi)=200$ MHz (i.e., $\omega_{q}-\omega_{b}\gg g$).
The other parameters are the same as in Fig.~\ref{fig2}, except
$\Delta_{a}/(2\pi)=100$ MHz and $|\Omega|/(2\pi) =19.8$ MHz. The
absorption spectrum shown in panel (a) explains the occurrence of
the optomechanical analog of two-color EIT. By contrast, the
spectra in the other panels exhibit only single transparency
windows. In particular, for far detunings shown in panels (d,e),
the EIT windows can roughly approximate the symmetric EIT window
for the standard optomechanical single-color EIT, when there is no
coupling between the mechanical resonator and the qubit, as shown
by the blue curve in Fig.~\ref{fig5}(a). Thus, it is seen how to
switch between the single and double transparency windows simply
by tuning the qubit transition frequency in or out of the
resonance with the mechanical-mode frequency.}\label{fig7}
\end{figure}
\begin{figure}
 \includegraphics[bb=20 115 570 720, width=8.6 cm,clip]{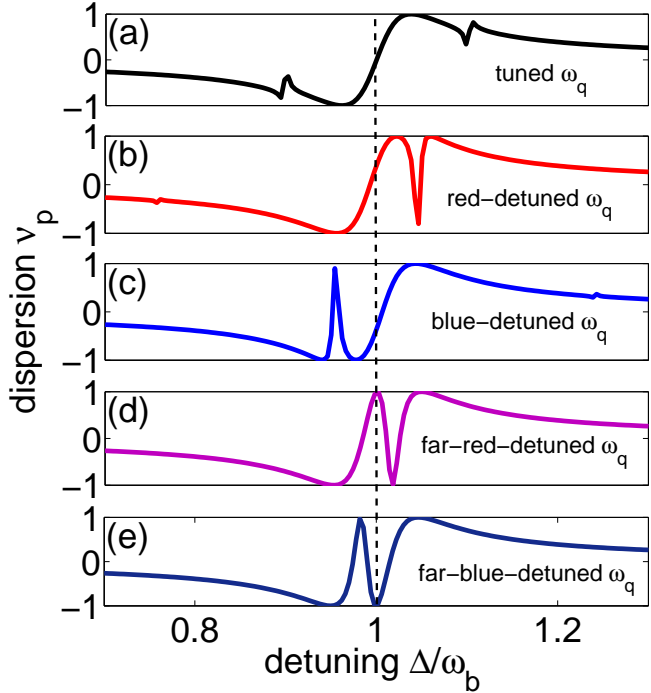}
\caption[]{(Color online) Same as in Fig.~\ref{fig7} but for the
dispersion spectra $\nu_p$ of the probe field.}\label{fig8}
\end{figure}
\begin{figure}
\includegraphics[bb=40 160 545 660, width=8.6 cm, clip]{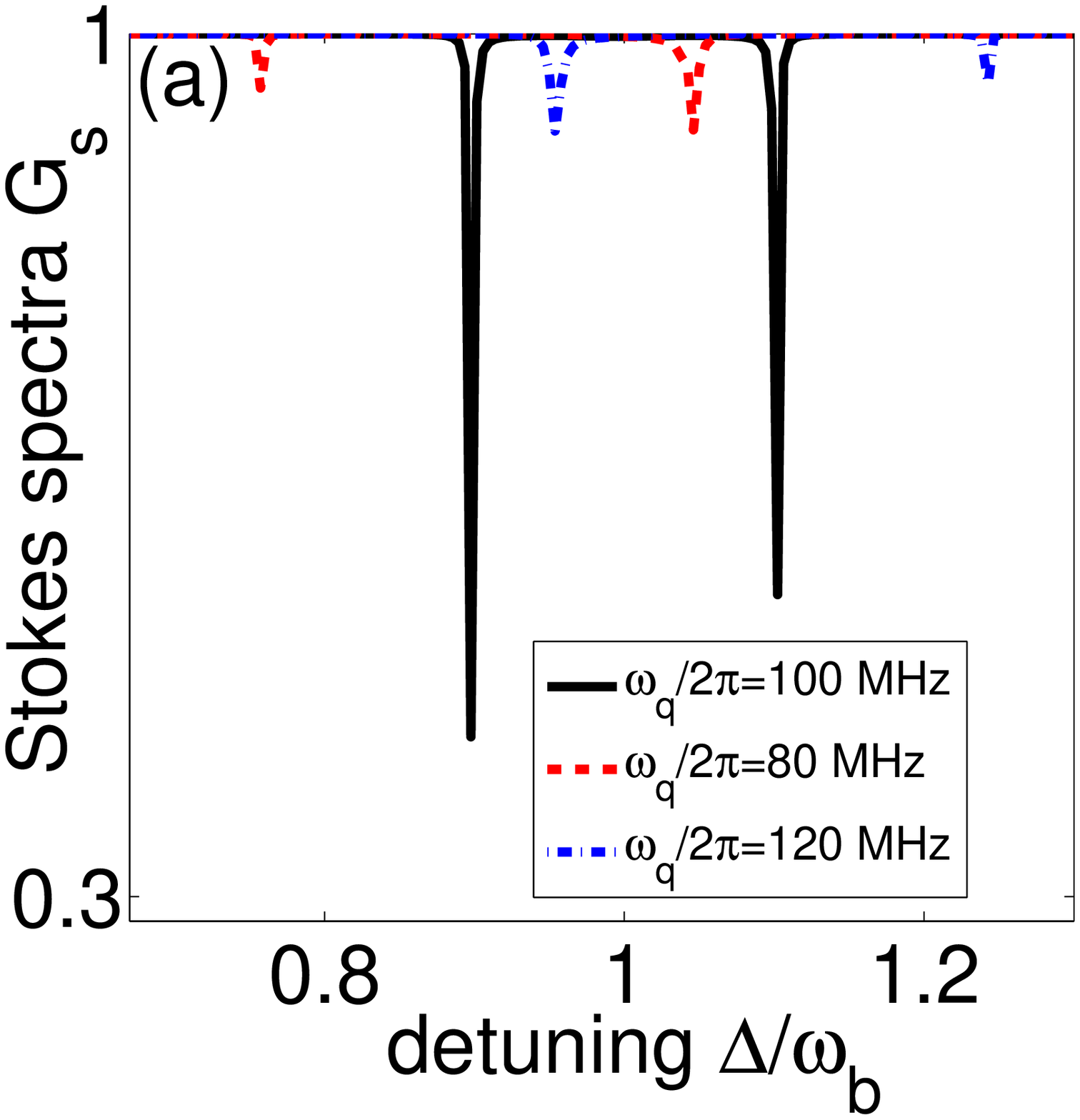}\\
\includegraphics[bb=-15 110 570 725, width=8.6 cm, clip]{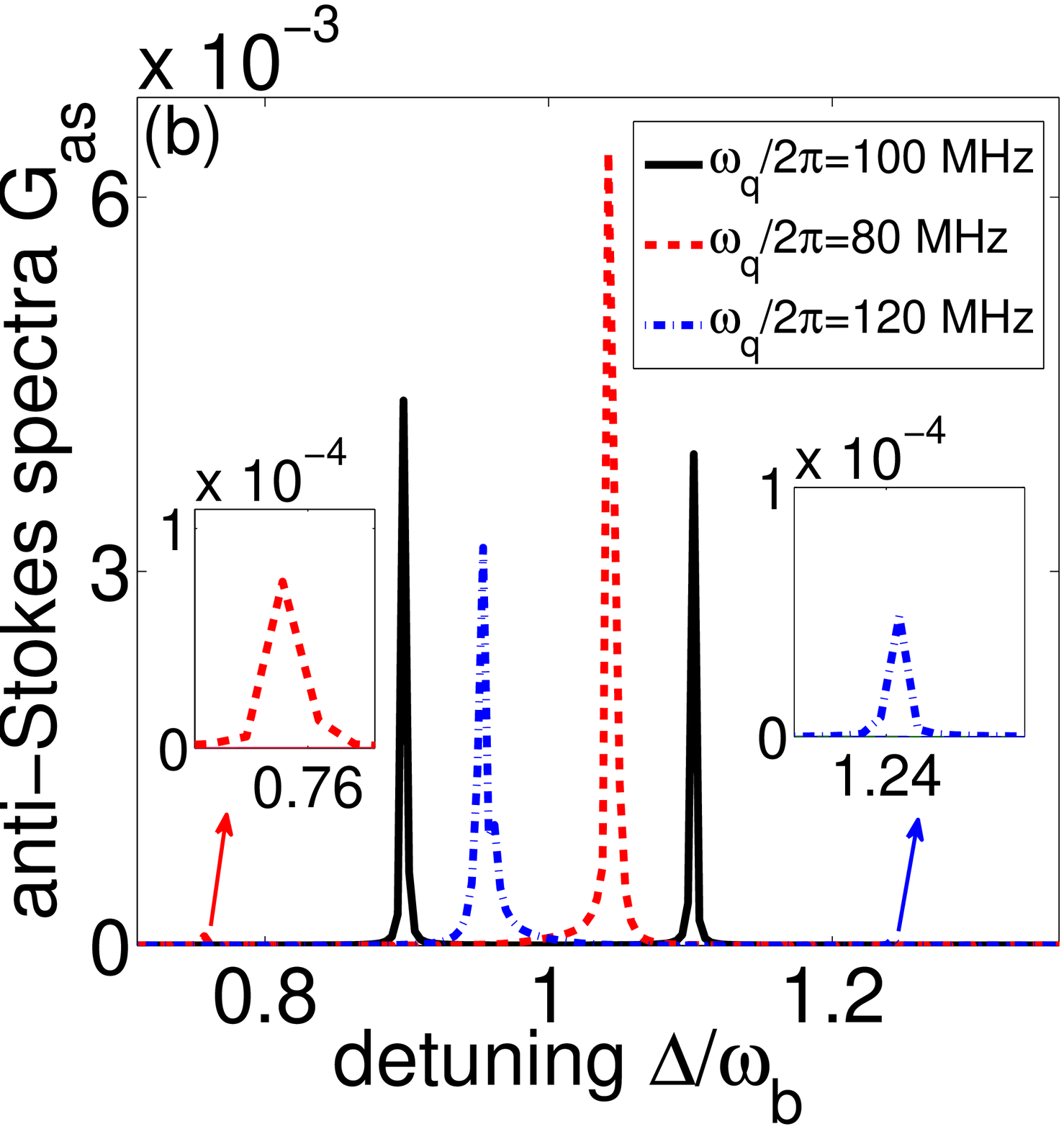}
\caption[]{(Color online) (a) The Stokes $G_{s}$ and (b) anti-Stokes
$G_{as}$ spectra of the output of the probe field versus the
detuning $\Delta=\omega_{p}-\omega_{d}$. In  each figure, three
curves correspond to different transition frequencies of the
qubit: (i) $\omega_{q}/(2\pi)=100$ MHz (black solid curve) being
in resonance with the mechanical-mode frequency $\omega_{b}$, as
shown in Figs.~\ref{fig7}(a) and~\ref{fig8}(a); (ii)
$\omega_{q}/(2\pi)=80$ MHz (red dashed curve) corresponding to the
case of the red-detuned $\omega_{q}$ shown in Figs.~\ref{fig7}(b)
and~\ref{fig8}(b); and (iii) $\omega_{q}/(2\pi)=120$ MHz (blue
dash-dotted curve) corresponding to the case of the blue-detuned
$\omega_{q}$ shown in Figs.~\ref{fig7}(c) and~\ref{fig8}(c). All
the parameters are the same as in the respective Figs.~\ref{fig7}
and~\ref{fig8}.}\label{fig9}
\end{figure}

In order to better understand the physical meaning of these
results, let us use the single-photon and single-phonon
excitations as an example to illustrate the nature of photon
transmission in this hybrid system (see Fig.~\ref{fig4}). The energy-level diagram for
the EIT analogue in optomechanical systems can be understood as in
Ref.~\cite{Weis}: the $\Lambda$-type three-level systems formed by
three states $|0_{a}, 0_{b}\rangle$, $|0_{a}, 1_{b}\rangle$, and
$|1_{a}, 0_{b}\rangle$. Here the subscripts $a$ and $b$ denote the
photon and phonon states, respectively. However, when a two-level
system is coupled to the mechanical resonator, the state $|0_{a},
1_{b}\rangle$ is split into two states $|0_{a},1_{b}+\rangle$ and
$|0_{a}, 1_{b}-\rangle$. Here $|1_{b}\pm\rangle$ denote the
dressed states~\cite{M-Orszag} formed by the single-phonon state
and the two-level system, e.g., $|1_{b}\pm\rangle=(|
1_{b},g\rangle\pm |0_{b},e\rangle)/\sqrt{2}$ for
$\omega_{b}=\omega_{q}$. This splitting of the single-phonon state
significantly affects the photon transmission if the detuning
between the cavity field and the driving field resonant with the
frequency of the mechanical resonator, i.e.,
$\Delta_{a}=\omega_{b}$. Clearly this splitting leads to two
transparency windows, which coincide well with the numerical
calculation shown in Fig.~\ref{fig5} and described below. The case
of multi-phonon excitations is very similar to the single-phonon
excitation, but the splitting width of the transparency windows
becomes wider. Moreover, the nonlinear coupling between the cavity
field and mechanical resonator makes the transmission spectrum
more complicated when the excitation numbers of the photon and
phonon are increased.

Figures~\ref{fig5}(a) and~\ref{fig5}(b) show, respectively, the
absorption and dispersion of the probe field for different values
of the coupling strength $g$ between the mechanical resonator and
the two-level system. These figures show a familiar transparency
window of the optomechanical system, which can occur when there is
no coupling of the two-level system to the mechanical resonator
(as shown by the blue solid curves in Fig.~\ref{fig5}). However,
two transparency windows can occur when the two-level system is
coupled to the mechanical resonator (as shown, e.g., by the red
dashed curves in Fig.~\ref{fig5}). The splitting of these two
transparency windows is equal to the splitting width $2g$ that
results from the Jaynes-Cummings coupling between the two-level
system and the mechanical mode. In Fig.~\ref{fig6}, the Stokes and
anti-Stokes power spectra are plotted as a function of the
frequency of the probe field. These spectra also show that the
two-level system changes the splitting width of the output spectra
at the Stokes and anti-Stokes frequencies.

In addition to the coupling strength $g$ between the two-level
system and the mechanical resonator, the transition frequency
$\omega_{q}$ of two-level system also affects the transmission of
the probe field, which will even more clearly show the main result
of our work. That is,  we find that the optomechanical analog of
two-color EIT and demonstrate that two-color EIT can be switched to the
standard single-color EIT.

In Fig.~\ref{fig7}, the absorption spectra of the probe field are
plotted as a function of the detuning $\Delta$ between the probe
and driving fields. Different panels of Fig.~\ref{fig7} show the
spectra for different values of the transition frequency of the
two-level system in comparison to the mechanical-mode frequency.
We observe in Fig.~\ref{fig7}(a) that there are two nearly
symmetric transparency windows [shown also by the red dashed curve
in Fig.~\ref{fig5}(a)] when the two-level system resonates with
the mechanical resonator. We refer to this effect as the
\emph{optomechanical analog of two-color EIT}. However, these two
transparency windows become asymmetric when the two-level system
is detuned from the mechanical resonator  frequency (as shown in
other panels of Fig.~\ref{fig7}). When the detuning
$|\omega_{q}-\omega_{b}|$ becomes much larger than the coupling
strength $g$, as in the cases shown in Figs.~\ref{fig7}(d,e),
these two transparency windows combine into one, almost symmetric,
window, which is very similar to that of the optomechanical
resonator without the qubit, as shown by the blue solid curve in
Fig.~\ref{fig5}(a). Thus, it is seen how to switch between one and
two transparency windows by changing the transition frequency of
the qubit, in particular how to approach the standard symmetric
EIT window in the far-detuning limits for
$|\omega_{q}-\omega_{b}|\gg g$. This change from two-color EIT to
single-color EIT, by detuning the qubit transition frequency
$\omega_{q}$ out of the resonance with the mechanical-mode
frequency $\omega_{b}$, can also be observed in other spectra. Our
examples include: (i) the dispersion spectra in Fig.~\ref{fig8},
which can be compared with Fig.~\ref{fig5}(b); (ii) the Stokes
spectra, i.e., the power spectra of the output at the Stokes
frequency, as shown in Fig.~\ref{fig9}(a) and to be compared with
Fig.~\ref{fig6}(a); and, analogously, (iii) the anti-Stokes
spectra presented in Fig.~\ref{fig9}(b). This figure can be
compared with Fig.~\ref{fig6}(b).

\section{Conclusions}\label{conclusion}

In summary, we have studied the transmission of a probe field
through an optomechanical system, consisting of a cavity and a
mechanical resonator with a two-level system, for simplicity
referred to as a qubit. The qubit might be an intrinsic defect
inside the mechanical resonator, a superconducting artificial
atom, or another two-level system. We assume that the mechanical
resonator is coupled to the qubit via the Jaynes-Cummings
interaction and to the cavity field via radiation pressure.

We find that the transmission of the probe field exhibits two
transparent windows when the two-level system is resonantly
coupled to the mechanical resonator. This is because the
interaction between the mechanical resonator and the two-level
system might result in two sets of coupling configurations between
the controlling field and the mechanical resonator. We consider
this effect as an optomechanical analog of two-color EIT (or
double EIT), in contrast to the standard optomechanical
single-color (or single-window) EIT exhibiting clear differences
in the probe-field spectra. Our examples include: the absorption,
dispersion, Stokes, and anti-Stokes spectra. We demonstrated how
to switch between one and two EIT windows by changing the
transition frequency of the qubit to be in or out of the resonance
with the frequency of the mechanical mode. These features might be
used to probe the low-frequency two-level fluctuations inside
solid-state systems by using a low-frequency mechanical resonator.
We note that the control of the transition frequency of the qubit
could be realized easier with an artificial two-level system
(e.g., a superconducting qubit) rather than with a natural
two-level defect.

In addition to optical switching, applications of the
optomechanical two-color EIT can include: the generation of
nonclassical states of microwave radiation and/or mechanical
resonator, nonlinear wave-mixing, cross-phase modulation,
wavelength conversion or photon blockade~\cite{HuiWang}, in
analogy to such applications of the standard optomechanical
single-color EIT.

We hope that our study, in particular, the finding of the
optomechanical analog of two-color EIT, which can be switched to
the standard single-color EIT, might provide a new method to
control the light transmission through optomechanical systems by
using a two-level system and to probe a low-frequency two-level
system by using a mechanical resonator. We also mention since the
dressed mechanical mode and the two-level system have the more
complicated energy structure when the mechanical mode is coupled
to the cavity field, thus, this hybrid system might exhibit more
phenomena as in the conventional atomic vapor EIT systems, for
example, the EIT in the multi-level atomic systems. These could be
explored in the future works~\cite{HuiWang}.

\section{Acknowledgement}
We thank Hui Jing and Jieqiao Liao for insightful and informative
discussions. Y.X.L. is supported by the National Natural Science
Foundation of China under Grant Nos. 61025022, 61328502, the
National Basic Research Program of China Grant No. 2014CB921401.
A.M. is supported by Grant No. DEC-2011/03/B/ST2/01903 of the
Polish National Science Centre. F.N. is partially supported by the
RIKEN iTHES Project, MURI Center for Dynamic Magneto-Optics,
JSPS-RFBR contract No. 12-02-92100, and Grant-in-Aid for
Scientific Research (S).

\appendix*

 \section{ Calculation of $A_{+}$ and $A_{-}$}\label{App}

From the discussions in Sec.~III, one can find the expressions of
$Z_{0}$, $Z_{+}$, $Z_{-}$ $L_{0}$, $L_{+}$, and $L_{-}$ in
Eqs.~(\ref{eq:23})--(\ref{eq:28}) up to first order in the
parameter $\varepsilon$ of the probe field by equating the
coefficients of the same order. Then the corresponding
coefficients are found to be
\begin{eqnarray}
\lambda_{2}&=&\frac{1}{D_{3}}\left[i g Z_{0}D_{1}-g\lambda_{1}B_{0}L^{\ast}_{0}\left(iD_{1}-g\lambda_{1} |B_{0}|^{2}\right)\right],\label{eq:78}\qquad\\
\lambda_{3}&=&i\frac{g\lambda_{1}B_{0}}{D_{3}}\left(ig\lambda_{1}|B_{0}|^{2}L_{0}+i gB_{0}  Z_{0} +D_{1}L_{0}\right),\label{eq:79}\\
\lambda_{4}&=&\frac{1}{D_{3}^{*}}\left[i g Z_{0}D_{2}+g\lambda^{\ast}_{1}B_{0}L^{\ast}_{0}\left(iD_{2}+ g\lambda^{\ast }_{1} |B_{0}|^{2}\right)\right],\label{eq:80}\\
\lambda_{5}&=&i\frac{g\lambda^{\ast}_{1}B_{0}}{D^{*}_{3}}\left(ig\lambda^{\ast}_{1}|B_{0}|^{2}L_{0}-i
gB_{0}  Z_{0} -D_{2}L_{0}\right),\qquad\label{eq:81}
\end{eqnarray}
where the parameters $D_{1}$, $D_{2}$, and $D_{3}$ are given by
\begin{eqnarray}
 D_{1}&=&\frac{\gamma_{q}}{2}-i\omega_{q}-ig\lambda_{1}|B_{0}|^{2}+i\Delta,\\ \label{eq:82}
 D_{2}&=&\frac{\gamma_{q}}{2}-i\omega_{q}+ig\lambda^{\ast}_{1}|B_{0}|^{2}-i\Delta,\\ \label{eq:83}
 D_{3}&=&\left(\frac{\gamma_{q}}{2}+i\Delta\right)^2-2ig\lambda_{1}|B_{0}|^{2}\left(\frac{\gamma_{q}}{2}+i\Delta\right)+\omega^{2}_{q}.\qquad \label{eq:84}
\end{eqnarray}
By substituting the expressions of $\langle b\rangle$, $\langle
a\rangle$, and $\langle \sigma_{-}\rangle $ into the equation of
motion for the average value of the operator $b$, we obtain the
expressions of $B_{0}$, $B_{+}$, and $B_{-}$ in
Eqs.~(\ref{eq:30})--(\ref{eq:32}). Here the coefficients
$\lambda_{6}$ and $\lambda_{7}$ are found as
\begin{eqnarray}
\lambda_{6}&=&\frac{-g\chi \lambda_{3}+i\chi D_{4}}{D_{4}D_{5}-g^{2}\lambda_{3}\lambda^{\ast}_{5}},\label{eq:88}\\
\lambda_{7}&=&\frac{-g\chi \lambda_{5}+i\chi
D^{\ast}_{5}}{D^{\ast}_{4}D^{\ast}_{5}-g^{2}\lambda^{\ast}_{3}\lambda_{5}}.\label{eq:89}
\end{eqnarray}
with
\begin{eqnarray}
D_{4}&=&\gamma_{b}-i\left(\omega_{b}-\Delta +g \lambda^{\ast}_{4}\right),\\
D_{5}&=&\gamma_{b}+i\left(\omega_{b}+\Delta+g \lambda_{2}\right).
\end{eqnarray}
Using similar steps as above, we obtain formulas for $A_{0}$,
$A_{-}$, and $A_{+}$ for the average value of the cavity field, up
to first order in the parameter $\varepsilon$ of the probe field
as given, respectively, by Eqs.~(\ref{eq:33})--(\ref{eq:35}) with
the parameters
\begin{eqnarray*}
 \lambda_{8}&=&\gamma_{a}+i\left[\Delta_{a}-\Delta -\chi B_{0}-\chi\left( \lambda^{\ast}_{6}+\lambda_{7}\right)|A_{0}|^{2}\right],\quad\\ \label{eq:91}
 \lambda_{9}&=&\gamma_{a}-i\left[\Delta_{a}+\Delta-\chi B^{\ast}_{0}-\chi\left( \lambda^{\ast}_{6}+\lambda_{7}\right)|A_{0}|^{2}\right],\quad\\ \label{eq:92}
 \lambda_{10}&=&\chi^{2}\left(\lambda^{\ast}_{6}+\lambda_{7}\right)^{2}|A_{0}|^{4}.\label{eq:93}
 \end{eqnarray*}
It is clear that $A_{0}$ represents the steady-state value of the
cavity field when the probe field is not applied to the cavity.
However, $A_{-}$ describes the linear response of the system to
the probe field, and $A_{+}$  describes the four-wave mixing of
the probe and driving fields.


\begin{thebibliography}{99}
\bibitem{PR1} M.P. Blencowe, Phys. Rep.~\textbf{395}, 159 (2004).



\bibitem{PR2}M. Poot and H. S. J. van der Zant, Phys. Rep.~\textbf{511}, 273 (2012).



\bibitem{ZLXiang}Z. L. Xiang, S. Ashhab, J. Q. You, and F. Nori, Rev. Mod. Phys.~\textbf{85}, 623 (2013).



\bibitem{Irish}E. K. Irish and K. Schwab, Phys. Rev. B~\textbf{68}, 155311 (2003).



\bibitem{amour}A. D. Armour, M. P. Blencowe, and K. C. Schwab,
Phys. Rev. Lett.~\textbf{88}, 148301 (2002).



\bibitem{Sornborger}A. T. Sornborger, A. N. Cleland, and M. R. Geller,
\pra~\textbf{70}, 052315 (2004).



\bibitem{Zhang}P. Zhang, Y. D. Wang, and C. P. Sun, \prl~\textbf{95}, 097204 (2005).



\bibitem{Xue} F. Xue, Y. D. Wang, C.P. Sun, H. Okamoto, H. Yamaguchi, and K. Semba, New J. Phys.~\textbf{9}, 35 (2007).



\bibitem{cleland}A. N. Cleland and M. R. Geller, Phys. Rev. Lett.~\textbf{93}, 070501 (2004);
M. R. Geller and A. N. Cleland, \pra~\textbf{71}, 032311 (2005).



\bibitem{tianT} L. Tian, Phys. Rev. B~\textbf{72}, 195411 (2005).



\bibitem{wei-liu} L. F. Wei, Y. X. Liu, C. P. Sun, and F. Nori, Phys. Rev. Lett.~\textbf{97}, 237201 (2006).



\bibitem{yuxiliu}Y. X. Liu, A. Miranowicz, Y. B. Gao, J. Bajer, C. P. Sun, and F. Nori,
\pra~\textbf{82}, 032101 (2010).



\bibitem{Roukes1}M. D. LaHaye, J. Suh, P. M. Echternach, K. C. Schwab, and M. L. Roukes, Nature (London)~\textbf{459}, 960 (2009).



\bibitem{feixue1}F. Xue, Y. X. Liu, C. P. Sun, and F. Nori, Phys. Rev. B~\textbf{76}, 064305 (2007).



\bibitem{feixue2}F. Xue, Y. D. Wang, Y. X. Liu, and F. Nori, Phys. Rev. B~\textbf{76}, 205302 (2007).



\bibitem{yongli}Y. Li, Y.-D. Wang, F. Xue, and C. Bruder, Phys. Rev. B~\textbf{78}, 134301 (2008).



\bibitem{T1}T. Rocheleau, T. Ndukum, C. Macklin, J. B. Hertzberg, A. A. Clerk, and K. C. Schwab, Nature (London)~\textbf{463}, 72 (2010).



\bibitem{T2}J. B. Hertzberg, T. Rocheleau, T. Ndukum, M. Savva, A. A. Clerk, and K. C. Schwab, Nat. Phys.~\textbf{6}, 213 (2010).



\bibitem{T3} F. Massel, T. T. Heikkil\"a, J.-M. Pirkkalainen, S. U. Cho, H. Saloniemi, P. J. Hakonen, and M. A. Sillanp\"a\"a,
Nature (London)~\textbf{480}, 351 (2011).



\bibitem{review1}T. J. Kippenberg and K. J. Vahala, Science~\textbf{321}, 1172 (2008).



\bibitem{review2}M. Aspelmeyer, S. Gr\"oblacher, K. Hammerer, and N. Kiesel, J. Opt. Soc. Am. B~\textbf{27}, A189 (2010).



\bibitem{review3}M. Aspelmeyer, P. Meystre, and K. Schwab, Phys. Today~\textbf{65}, 29 (2012).



\bibitem{review4}M. Aspelmeyer, T. J. Kippenberg, and F. Marquardt,  arXiv:1303.0733.



\bibitem{NV1}O. Arcizet, V. Jacques, A. Siria, P. Poncharal, P. Vincent, and S. Seidelin, Nat. Phys.~\textbf{7}, 879 (2011).



\bibitem{NV2}S. Kolkowitz, A. C. B. Jayich, Q. P. Unterreithmeier, S. D. Bennett, P. Rabl, J. G. E. Harris, and M.D. Lukin, Science~\textbf{335}, 1603 (2012).



\bibitem{NV3}S. D. Bennett, N. Y. Yao, J. Otterbach, P. Zoller, P. Rabl, and M. D. Lukin, Phys. Rev. Lett.~\textbf{110},
156402 (2013).



\bibitem{Rugar} D. Rugar, R. Budakian, H. J. Mamin, and B. W. Chui, Nature (London)~\textbf{430}, 329 (2004).



\bibitem{Tdefects}L. Tian, Phys. Rev. B~\textbf{84}, 035417 (2011).



\bibitem{Grabovskij}
G. J. Grabovskij, T. Peichel, J. Lisenfeld, G. Weiss, and A. V.
Ustinov, Science~\textbf{338}, 232 (2012).



\bibitem{phononblockade}N. Didier, S. Pugnetti, Y. M. Blanter, and R. Fazio, Phys. Rev. B~\textbf{84}, 054503 (2011).



\bibitem{Q1}A. D. O'Connell, M. Hofheinz, M. Ansmann, R. C. Bialczak, M. Lenander, E. Lucero, M. Neeley, D. Sank, H. Wang, M.
Weides, J. Wenner, J. M. Martinis, and A. N. Cleland, Nature
(London)~\textbf{464}, 697 (2010).



\bibitem{Q2}J. D. Teufel, T. Donner, D. Li, J. W. Harlow, M. S. Allman, K. Cicak, A. J. Sirois, J. D.
Whittaker, K. W. Lehnert, and R. W. Simmonds, Nature
(London)~\textbf{475},  359 (2011).



\bibitem{Q3}J. Chan, T. P. M. Alegre, A. H. Safavi-Naeini, J. T. Hill, A. Krause, S. Groblacher, M. Aspelmeyer, and O. Painter,
Nature (London)~\textbf{478}, 89 (2011).



\bibitem{Q4}E. Verhagen, S. Deleglise, S. Weis, A. Schliesser, and T. J. Kippenberg, Nature (London)~\textbf{482}, 63 (2012).



\bibitem{sun}C. P. Sun, L. F. Wei, Y. X. Liu, and F. Nori, \pra~\textbf{73}, 022318
(2006).



\bibitem{Pirkkalainen}J.-M. Pirkkalainen, S. U. Cho, J. Li, G. S. Paraoanu, P. J. Hakonen, and M. A. Sillanp\"{a}\"{a}, Nature (London)~ \textbf{494}, 211(2013).


\bibitem{gaoming}M. Gao, Y. X. Liu, and X. B. Wang, \pra~\textbf{83}, 022309 (2011).



\bibitem{tian1}L. Tian and H. Wang, \pra~\textbf{82}, 053806 (2010).



\bibitem{ying-dan}Y. D. Wang and A. A. Clerk, Phys. Rev. Lett.~\textbf{108}, 153603 (2012).



\bibitem{tian}L. Tian, Phys. Rev. Lett.~\textbf{108}, 153604 (2012).



\bibitem{Hill}J. T. Hill, A. H. Safavi-Naeini, J. Chan, and O. Painter,
Nat. Commun.~\textbf{3}, 1196 (2012).



\bibitem{Hailing}C. H. Dong, V. Fiore, M. C. Kuzyk, and H. Wang, Science~\textbf{338}, 1609 (2012).



\bibitem{cleland2013}J. Bochmann, A. Vainsencher, D. Awschalom, and A. N. Cleland, Nat. Phys.~\textbf{9}, 712 (2013).



\bibitem{Lu13}
X.Y. L\"u, W.M. Zhang, S. Ashhab, Y. Wu, and F. Nori, Sc. Reports
\textbf{3}, 2943 (2013).



\bibitem{cat}
B. Yurke and D. Stoler, \prl \textbf{57,} 13 (1986); P. Tombesi
and A. Mecozzi, J. Opt. Soc. Am. B \textbf{4,} 1700 (1987).



\bibitem{kitten}
A. Miranowicz, R. Tana\'s, and S. Kielich, Quantum Opt.
\textbf{2}, 253 (1990); R. Tana\'s, Ts. Gantsog, A. Miranowicz,
and S. Kielich, \josab  \textbf{8}, 1576 (1991).



\bibitem{blockade2}A. Nunnenkamp, K. Borkje, and S. M. Girvin, Phys. Rev. Lett.~\textbf{107}, 063602 (2011).



\bibitem{binghe}B. He, \pra~\textbf{85}, 063820 (2012).



\bibitem{JQLiao1}J. Q. Liao, H. K. Cheung, and C. K. Law, \pra~\textbf{85}, 025803 (2012).



\bibitem{JQLiao2}J. Q. Liao and F. Nori, \pra~\textbf{88}, 023853 (2013).



\bibitem{xunwei}X. W. Xu, Y. J. Li, and Y. X. Liu, \pra~\textbf{87}, 025803 (2013).



\bibitem{germany}A. Kronwald, M. Ludwig, and F. Marquardt, \pra~\textbf{87}, 013847 (2013).



\bibitem{strong1}S. Gupta, K. L. Moore, K. W. Murch, and D. M. Stamper-Kurn, Phys. Rev. Lett.~\textbf{99}, 213601 (2007).



\bibitem{strong2}M. Eichenfield, J. Chan, R. M. Camacho, K. J. Vahala, and O. Painter, Nature (London)~\textbf{462}, 78 (2009).



\bibitem{strong3}J. D. Teufel, D. Li, M. S. Allman, K. Cicak, A. J. Sirois, J. D. Whittaker, and R. W. Simmonds, Nature (London)~\textbf{471}, 204 (2011).



\bibitem{strong4}J. C. Sankey, C. Yang, B. M. Zwickl, A. M. Jayich, and  J. G. E. Harris, Nat. Phys.~\textbf{6}, 707 (2010).



\bibitem{xu-phonon1}X. W. Xu, H. Wang, J. Zhang, and Y. X. Liu,  \pra~\textbf{88}, 063819 (2013).




\bibitem{xu-phonon2}X. W. Xu, Y. J. Zhao, and Y. X. Liu, \pra~\textbf{88}, 022325 (2013).




\bibitem{Agarwal} G. S. Agarwal and S. Huang, \pra~\textbf{81}, 041803(R) (2010).



\bibitem{Weis} S. Weis, R. Riviere, S. Deleglise, E. Gavartin,
O. Arcizet, A. Schliesser, and T. J. Kippenberg,
Science~\textbf{330}, 1520 (2010).



\bibitem{SafaviNaeini} A. H. Safavi-Naeini, T. P. Mayer Alegre,
J. Chan, M. Eichenfield, M. Winger, Q. Lin, J. T. Hill, D. E.
Chang, and O. Painter, Nature (London)~\textbf{472}, 69 (2011).



\bibitem{yingwu}H. Xiong, L.-G. Si, A.-S. Zheng, X. Yang, and Y. Wu,
\pra~\textbf{86}, 013815 (2012).




\bibitem {Harris1} K.-J. Boiler, A. Imamoglu, and S. E. Harris,  \prl ~\textbf{66}, 2593 (1991).



\bibitem {Harris2}S. E. Harris, Phys. Today ~\textbf{50}, No. 7, 36 (1997).



\bibitem {Peng} B. Peng, S. K. Ozdemir, W. Chen, F. Nori, and L. Yang,  arXiv:1404.5941v1 (2014).



\bibitem{Ian} H. Ian, Z. R. Gong, Y. X. Liu, C. P. Sun, and F. Nori,
\pra~\textbf{78}, 013824 (2008).




\bibitem{yuechang}Y. Chang, T. Shi, Y. X. Liu, C. P. Sun, and F. Nori, \pra~\textbf{83}, 063826 (2011).



\bibitem{huijing} H. Jing and X. Zhao, and L. F. Buchmann \pra~\textbf{86}, 065801
(2012); H. Jing, D. S. Goldbaum, L. Buchmann, and P. Meystre,
\prl~\textbf{106}, 223601 (2011).



\bibitem{gaoxiangli}L. H. Sun, G. X. Li, and Z. Ficek, \pra~\textbf{85}, 022327 (2012).




\bibitem{huiwang}H. Wang, H. C. Sun, J. Zhang, and Y. X. Liu, Sci. China-Phys. Mech. Astron.~\textbf{55}, 2264 (2012).



\bibitem{hybrid1} D. Breyer and M. Bienert, \pra~\textbf{86}, 053819 (2012).



\bibitem{hybrid7} W. Z. Jia and Z. D. Wang,  \pra~\textbf{88}, 063821 (2013).



\bibitem{Ian10} H. Ian, Y. X. Liu, and F. Nori, \pra~\textbf{81}, 063823 (2010).



\bibitem{Sun13} H.C. Sun, Y. X. Liu, J. Q. You, E. Il'ichev, and F. Nori, \pra~\textbf{89},  063822 (2014).



\bibitem{Bsanders}P. M. Anisimov, J. P. Dowling, and B. C. Sanders,
Phys. Rev. Lett.~\textbf{107}, 163604 (2011).



\bibitem{Phillips} W. A. Phillips, J. Low Temp. Phys.~\textbf{7}, 351 (1972).



\bibitem{Mohanty} G. Zolfagharkhani,  A. Gaidarzhy, Seung-Bo Shim, R. L. Badzey, and P. Mohanty, \prb ~\textbf{72}, 224101 (2005).



\bibitem{Arcizet} O. Arcizet, R. Rivi¨¨re, A. Schliesser, G. Anetsberger, and T. J. Kippenberg, \pra~\textbf{80}, 021803 (2009).




\bibitem{hybrid2} T. Ramos, V. Sudhir, K. Stannigel, P. Zoller, and T. J. Kippenberg,
Phys. Rev. Lett.~\textbf{110}, 193602 (2013).



\bibitem{Hammerer} K. Hammerer, M. Wallquist, C. Genes, M. Ludwig, F. Marquardt, P. Treutlein,
P. Zoller, J. Ye, and H. J. Kimble, \prl \textbf{103}, 063005
(2009).




\bibitem{You}J.Q. You, F. Nori, Physics Today \textbf{58} (11), 42 (2005); Nature (London) \textbf{474}, 589 (2011).


\bibitem{2colorEIT}
 H. Yan, K.-Y. Liao, J.-F. Li, Y.-X. Du, Z.-M. Zhang, and S.-L. Zhu, \pra~\textbf{87}, 055401 (2013);
 Y. Liu, J. Wu, D. Ding, B. Shi, and G. Guo, New J. Phys.~\textbf{14}, 073047 (2012);
 G.-Q. Yang, P. Xu, J. Wang, Y. Zhu, and M. S. Zhan, \pra~\textbf{82}, 045804 (2010);
 S. A. Moiseev and B. S. Ham, \pra~\textbf{73}, 033812 (2006);
 J. Wang, Y. Zhu, K. J. Jiang, and M. S. Zhan, \pra~\textbf{68}, 063810 (2003).



\bibitem{He07}
L. He, Y.X . Liu, S. Yi, C. P. Sun, and F. Nori, \pra~\textbf{75},
063818 (2007).



\bibitem{NonlinearOptics}R. W. Boyd, \textit{Nonlinear Optics} (Academic Press, New York, 2010).



\bibitem{mancini}S. Mancini and P. Tombesi, \pra~\textbf{49}, 4055 (1994).



\bibitem{walls} D. F. Walls and G. J. Milburn, \textit{Quantum
Optics} (Springer, Berlin, 1994).




\bibitem{Huang10}
S. Huang and G. S. Agarwal, \pra~\textbf{81}, 053810 (2010).



\bibitem {Shu1}S. Li, X. Yang, X. Cao, C. Xie, and H. Wang, J. Phys. B~\textbf{40}, 3211 (2007).



\bibitem{Goren}C. Goren, A. D. Wilson-Gordon, M. Rosenbluh, and H. Friedmann, \pra~\textbf{69}, 063802 (2004).



\bibitem {Shu2} S. Li, X. Yang, X. Cao, C. Zhang, C. Xie, and H. Wang, \prl~\textbf{101}, 073602 (2008).



\bibitem{M-Orszag} M. Orszag, \textit{Quantum Optics} (Springer, Berlin, 2000).


\bibitem{HuiWang}
H. Wang \emph{et al.}, in preparation (2014).



\end{thebibliography}
\end{document}